\documentclass[aps,prd,preprint,showkeys,superscriptaddress]{revtex4-1}

\usepackage{xcolor}
\usepackage{ulem}

\usepackage{amsmath}
\usepackage{graphicx}

\newcommand{\Slash}[1]{{\ooalign{\hfil/\hfil\crcr$#1$}}}
\newcommand{\ovl}[1]{\overline{#1}}

\newcommand{\eqn}[1]{(\ref{#1})}
\newcommand{\nn}{\nonumber}

\def\simge{\mathrel{%
       \rlap{\raise 0.511ex \hbox{$>$}}{\lower 0.511ex \hbox{$\sim$}}}}
\def\simle{\mathrel{
       \rlap{\raise 0.511ex \hbox{$<$}}{\lower 0.511ex \hbox{$\sim$}}}}

\begin{document} 

\title{Topological susceptibility in finite temperature (2+1)-flavor QCD using gradient
flow}

\preprint{UTHEP-697, UTCCS-P-93, KYUSHU-HET-172}

\author{Yusuke Taniguchi}
\email{tanigchi@het.ph.tsukuba.ac.jp}
\affiliation{Center for Computational Sciences (CCS), University of Tsukuba,
Tsukuba, Ibaraki 305-8577, Japan}
\author{Kazuyuki Kanaya}
\email{kanaya@ccs.tsukuba.ac.jp}
\affiliation{Center for Integrated Research in Fundamental Science and
Engineering (CiRfSE), University of Tsukuba, Tsukuba, Ibaraki 305-8571, Japan}
\author{Hiroshi Suzuki}
\email{hsuzuki@phys.kyushu-u.ac.jp}
\affiliation{Department of Physics, Kyushu University, 744 Motooka, Fukuoka
819-0395, Japan}
\author{Takashi Umeda}
\email{tumeda@hiroshima-u.ac.jp}
\affiliation{Graduate School of Education, Hiroshima University,
Higashihiroshima, Hiroshima 739-8524, Japan}
\collaboration{WHOT-QCD Collaboration}
\noaffiliation

\date{\today}

\begin{abstract}
We compute the topological charge and its susceptibility in finite temperature (2+1)-flavor QCD on the lattice applying a gradient flow method. 
With the Iwasaki gauge action and nonperturbatively $O(a)$-improved Wilson quarks, we perform simulations
on a fine lattice with~$a\simeq0.07\,\mathrm{fm}$ at a heavy $u$, $d$ quark mass with $m_\pi/m_\rho\simeq0.63$ but approximately physical $s$ quark mass with $m_{\eta_{ss}}/m_\phi\simeq0.74$.
In a  temperature range from~$T\simeq174\,\mathrm{MeV}$  ($N_t=16$) to $697\,\mathrm{MeV}$ ($N_t=4$), we study two topics on the topological susceptibility.
One is a comparison of gluonic and fermionic definitions of the topological susceptibility.
Because the two definitions are related by chiral Ward-Takahashi identities, their equivalence is
not trivial for lattice quarks which violate the chiral symmetry explicitly at finite lattice spacings.
The gradient flow method enables us to compute them without being bothered by the chiral violation. 
We find a good agreement between the two definitions with Wilson quarks.
The other is a comparison with a prediction of the dilute instanton gas approximation, which is relevant in a study of axions as a candidate of the dark matter in the evolution of the Universe.
We find that the topological susceptibility shows a decrease in $T$ which is consistent with the predicted 
$\chi_\mathrm{t}(T) \propto (T/T_\mathrm{pc})^{-8}$ for three-flavor QCD even at low temperature $T_\mathrm{pc} < T\simle1.5\,T_\mathrm{pc}$.
\end{abstract}

\maketitle

\section{Introduction}
\label{sec:intro}

The axion is introduced into QCD to solve the strong {\it CP} problem through the Peccei-Quinn mechanism \cite{Peccei:1977hh}.
Simultaneously, the axion is a candidate of the cold dark matter
where the temperature dependence of its mass plays an important role in the estimation of its cosmic abundance \cite{Preskill,Abbott,Dine}.
The axion mass squared is proportional to the topological susceptibility~$\chi_\mathrm{t}$. 
The temperature dependence of $\chi_\mathrm{t}$ is predicted by the dilute instanton gas approximation (DIGA) \cite{tHooft:1976snw}
to be $\chi_\mathrm{t}(T)\propto (T/T_\mathrm{pc})^{-8}$ in a high temperature limit for three flavors \cite{Gross:1980br}, where $T_\mathrm{pc}$ is the pseudocritical temperature.
Recently, $\chi_\mathrm{t}(T)$ is studied in lattice QCD
in the quenched approximation \cite{Berkowitz:2015aua,Kitano:2015fla,Borsanyi:2015cka}
and with (2+1)-flavors \cite{Bonati:2015vqz,Petreczky:2016vrs} 
or (2+1+1)-flavors~\cite{Borsanyi:2016ksw} of staggered quarks.
In~Ref.~\cite{Bonati:2015vqz}, the decreasing behavior is found to be much slower than DIGA, 
while in Ref.~\cite{Petreczky:2016vrs}, the power is found to be consistent with DIGA
above $1.5\,T_\mathrm{pc}$ but is a bit more moderate for $T_\mathrm{pc} < T\simle1.5\,T_\mathrm{pc}$.
In this paper, we study this issue in $N_{f}=$2+1 lattice QCD with improved Wilson quark action 
based on the gradient flow \cite{
Narayanan:2006rf,Luscher:2009eq,Luscher:2010iy,Luscher:2011bx,Luscher:2013cpa,reviewLattice}
and calculate the temperature dependence of topological charge and its susceptibility in the range $T\simeq174$-$697\,\mathrm{MeV} \simeq (0.92\mathrm{-}3.67) T_\mathrm{pc}$.

A problem in the calculation of topological charge on the lattice is the UV singularities in composite operators, which becomes acute when chiral symmetry is broken explicitly.
In particular, $\chi_\mathrm{t}$ defined in terms of gauge fields (``gluonic definition'') and that in terms of quark fields (``fermionic definition''), which are equivalent in the continuum theory or with Ginsparg-Wilson lattice quarks~\cite{Giusti:2004qd}, are largely discrepant with more conventional nonchiral lattice quarks.
For example, a recent study with improved staggered quarks reports more than 100 times larger gluonic $\chi_\mathrm{t}$ than fermionic one at $T\sim1.5\,T_\mathrm{pc}$ on finite lattices \cite{Petreczky:2016vrs}.
Much efforts have been dedicated to avoid the singular behavior \cite{Luscher:2004fu,Giusti:2008vb,Luscher:2010ik}.
We solve the above problem by making use of a UV divergence-free property of the gradient flow~\cite{Narayanan:2006rf,Luscher:2009eq,Luscher:2010iy,Luscher:2011bx,Luscher:2013cpa}.
This is an extension of our previous study on energy-momentum tensor and chiral condensate \cite{ourEOSpaper} using the method of Refs.~\cite{Suzuki:2013gza,Makino:2014taa,Hieda:2016lly}.

The gradient flow we adopt is described in Ref.~\cite{ourEOSpaper}.
The gauge field is flowed with fictitious time $t$ as \cite{Luscher:2010iy}
\begin{equation}
   \partial_tB_\mu(t,x)=D_\nu G_{\nu\mu}(t,x),\quad
   B_\mu(t=0,x)=A_\mu(x),
\label{eq:(1.1)}
\end{equation}
where 
\begin{eqnarray}
&&G_{\mu\nu}\equiv
\partial_\mu B_\nu-\partial_\nu B_\mu
+[B_\mu,B_\nu],
\label{eq:(1.2)}
\\
&&D_\nu G_{\nu\mu}\equiv
\partial_\nu G_{\nu\mu}+[B_\nu,G_{\nu\mu}].
\end{eqnarray}
The gradient flow for quark fields is given by \cite{Luscher:2013cpa}
\begin{eqnarray}
&&
\partial_t\chi_f(t,x)=\Delta\chi_f(t,x),
\quad
   \chi_f(t=0,x)=\psi_f(x),
\label{eq:(1.14)}
\\&&
\partial_t\Bar{\chi}_f(t,x)
   =\Bar{\chi}_f(t,x)\overleftarrow{\Delta},
   \quad
   \Bar{\chi}_f(t=0,x)=\Bar{\psi}_f(x),
\label{eq:(1.15)}
\end{eqnarray}
where $f=u$, $d$ and $s$, with
\begin{eqnarray}
&&
\Delta\chi_f\equiv D_\mu D_\mu\chi_f,
\quad
D_\mu\chi_f\equiv\left[\partial_\mu+B_\mu\right]\chi_f,
\\&&
\Bar{\chi}_f\overleftarrow{\Delta}
   \equiv\Bar{\chi}_f\overleftarrow{D}_\mu\overleftarrow{D}_\mu,
\quad
   \Bar{\chi}_f\overleftarrow{D}_\mu
   \equiv\Bar{\chi}_f\left[\overleftarrow{\partial}_\mu-B_\mu\right].
\end{eqnarray}

The flowed fields can be viewed as smeared fields over a range of about $\sqrt{8t}$ in four dimensions.
Operators constructed with the flowed fields are shown to be free from UV divergence
when multiplied with an appropriate wave function renormalization factor to the quark fields~\cite{Luscher:2011bx,Luscher:2013cpa}.
We can thus consider the flowed operators as renormalized operators in a new renormalization scheme with the scale $\sqrt{8t}$.

\section{Simulation parameters}
\label{sec:parameters}

Measurements are performed on $N_f=2+1$ QCD configurations generated for Refs.~\cite{Ishikawa:2007nn,Umeda:2012er}
adopting a nonperturbatively $O(a)$-improved Wilson quark action and the renormalization group-improved Iwasaki gauge action~\cite{Iwasaki:2011np}. 
Our gauge coupling constant $\beta=2.05$ corresponds to the lattice spacing~$a=0.0701(29)\,\mathrm{fm}$ ($1/a\simeq2.79\,\mathrm{GeV}$).
The hopping parameters $\kappa_u=\kappa_d\equiv\kappa_{ud}=0.1356$ and $\kappa_s=0.1351$ correspond to heavy $u$ and $d$ quarks, $m_\pi/m_\rho\simeq0.63$, and
almost physical $s$ quark, $m_{\eta_{ss}}/m_\phi\simeq0.74$. The bare PCAC
quark masses are $a\,m_{ud}=0.02105(17)$ and $a\,m_s=0.03524(26)$.
With the fixed-scale approach~\cite{Levkova:2006gn,Umeda:2008bd}, the temperature $T=1/(aN_t)$ is varied by changing the temporal lattice size $N_{t}$.
We adopt $N_t=16$, 14, 12, 10, 8, 6, and 4, which correspond to $T\simeq 174$, 199, 232, 279, 348, and 697 MeV, respectively
($T/T_\mathrm{pc}\simeq0.92$, 1.05, 1.22, $\cdots$ 3.67, assuming 
the pseudocritical temperature of $T_\mathrm{pc} \sim 190$ MeV~\cite{Umeda:2012er}).
See Table \ref{table:parameters} for temperature and number of configurations at each $N_{t}$.
Spatial box size is $N_s^3=32^3$ for finite temperature and $28^3$ for zero temperature.
To avoid unphysical effects due to overlapped smearing, 
we require
\begin{equation}
t \le t_{1/2} \equiv \frac{1}{8}\left[\min\left(N_t/2,N_s/2\right)\right]^2 .
\label{eq:t-half}
\end{equation}
Our study of the energy-momentum tensor and chiral condensate on these configurations suggests that our lattices are sufficiently fine 
but the lattices with $N_t\simle8$ suffer from small-$N_t$ lattice artifacts~\cite{ourEOSpaper}.

The differential equations for the gradient flow are solved 
by the third-order Runge-Kutta method~\cite{Luscher:2010iy,Luscher:2013cpa} with the step size of~$\epsilon=0.02$. 
Quark observables are evaluated with the noisy estimator method~\cite{ourEOSpaper}. 
The number of noise vectors is 20 for each color.
The statistical errors are estimated by a jackknife analysis with bin size of 300 in Monte Carlo time as determined from the autocorrelation.

\begin{table}[htb]
\centering
\begin{tabular}{cccccc}
 $T$ (MeV) & $T/T_{\mathrm{pc}}$ & $N_t$ & $t_{1/2}$ & Number of confs.\\
\hline
 $0$   & $0$    & $56$ &$24.5$ & $650$ \\
 $174$ & $0.92$ & $16$ &$8$ & $1440$ \\
 $199$ & $1.05$ & $14$ &$6.125$ & $1270$ \\
 $232$ & $1.22$ & $12$ &$4.5$ & $1290$ \\
 $279$ & $1.47$ & $10$ &$3.125$ & $780$ \\
 $348$ & $1.83$ &  $8$ &$2$ & $510$ \\
 $464$ & $2.44$ &  $6$ &$1.125$ & $500$ \\
 $697$ & $3.67$ &  $4$ &$0.5$ & $700$ \\
\hline
\end{tabular}
\caption{Parameters for the numerical simulation: Temperature in MeV,
$T/T_{\mathrm{pc}}$ assuming $T_{\mathrm{pc}}=190$ MeV, $N_t$,
 $t_{1/2}$, and number of configurations.
 Configurations are stored every five trajectories ($\tau=5$) at
 finite temperature and every $10$~trajectories at zero temperature.
 The bare coupling and the hopping parameter is set to $\beta=2.05$,
 $\kappa_{ud}=0.1356$, and $\kappa_s=0.1351$ for every~$N_t$. Spatial box
 size is $32^3$ for~$T>0$ and $28^3$ for~$T=0$.}
\label{table:parameters}
\end{table}%

\section{Gluonic definition}
\label{sec:gluonic}

The most popular definition of the topological charge is to use the gauge field strength $F\tilde{F}$
accompanied with a cooling step \cite{GarciaPerez:1998ru,Bonati:2014tqa,Namekawa:2015wua,Alexandrou:2015yba}.
The gradient flow provides us with a cooling procedure \cite{Bonati:2014tqa} and a renormalization procedure simultaneously. 
Let us define the topological charge density by the flowed gauge field
as~\cite{Luscher:2010iy}
\begin{equation}
   q(t,x)=\frac{1}{64\pi^2}
   \epsilon_{\mu\nu\rho\sigma}G_{\mu\nu}^a(t,x)G_{\rho\sigma}^a(t,x),
\quad
\epsilon_{0123}=1,
\label{eq:(1.4)}
\end{equation}
and the topological charge as $Q(t)=\int d^4x\,q(t,x)$.
There are several alternative choices of lattice operators
for the quadratic term of the field strength tensor $G_{\mu\nu}$.
In this study, we adopt the tree-level $O(a^4)$-improved field strength squared
by combing the clover operator with four plaquette Wilson loops 
and that with four $1\times2$ rectangle Wilson loops~\cite{AliKhan:2001ym}.
The normalization of the topological charge thus defined is shown to be consistent with the Ward-Takahashi (WT) relation associated with the flavor singlet chiral symmetry~\cite{Ce:2015qha,Hieda:2016lly}
and the operator $Q(t)$ is independent of the flow time in the continuum limit~\cite{Luscher-talk,Ce:2015qha}.

\begin{figure}[tbh]
 \begin{center}
  \includegraphics[width=7cm]{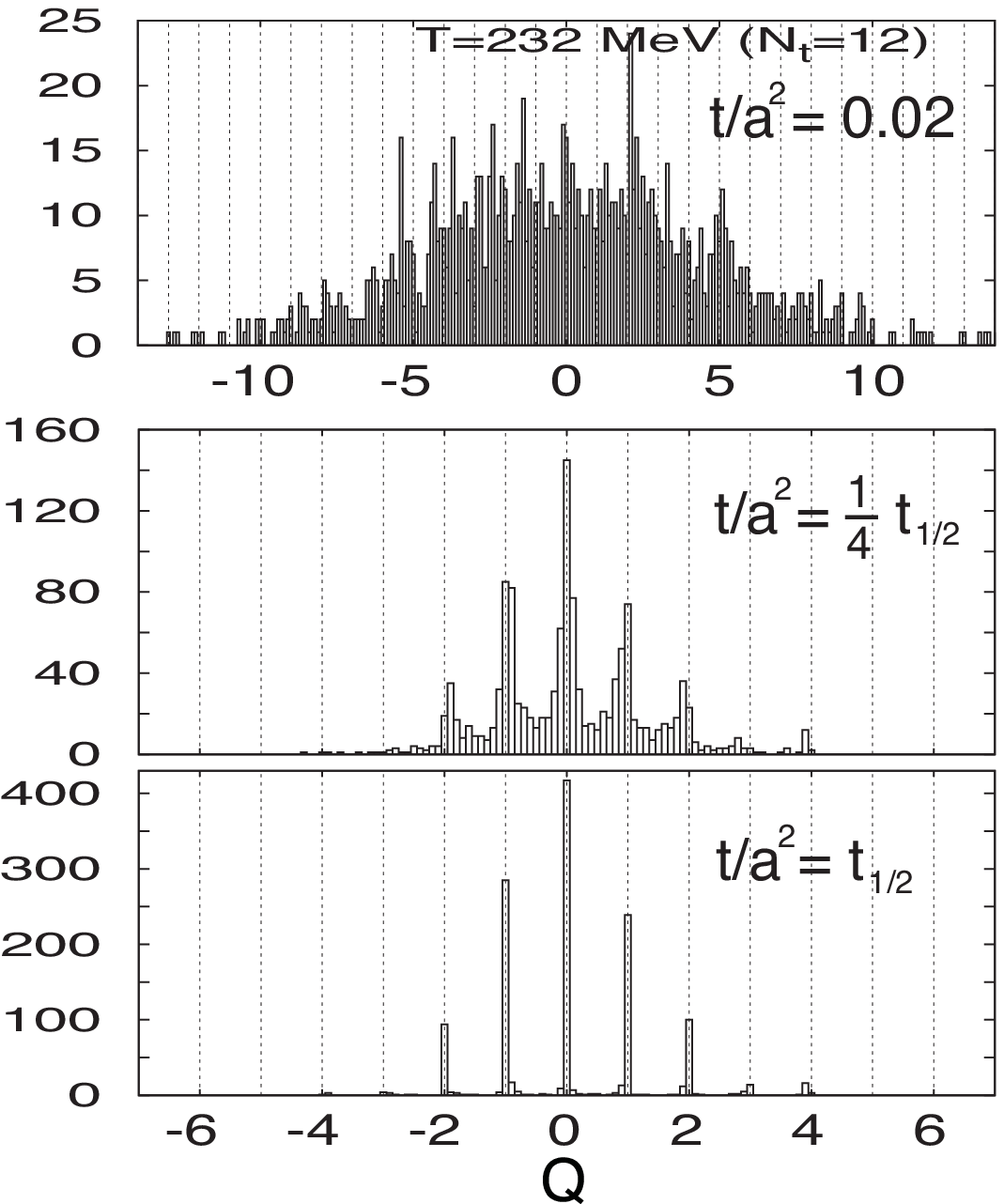}
  \hspace{12mm}
  \includegraphics[width=7cm]{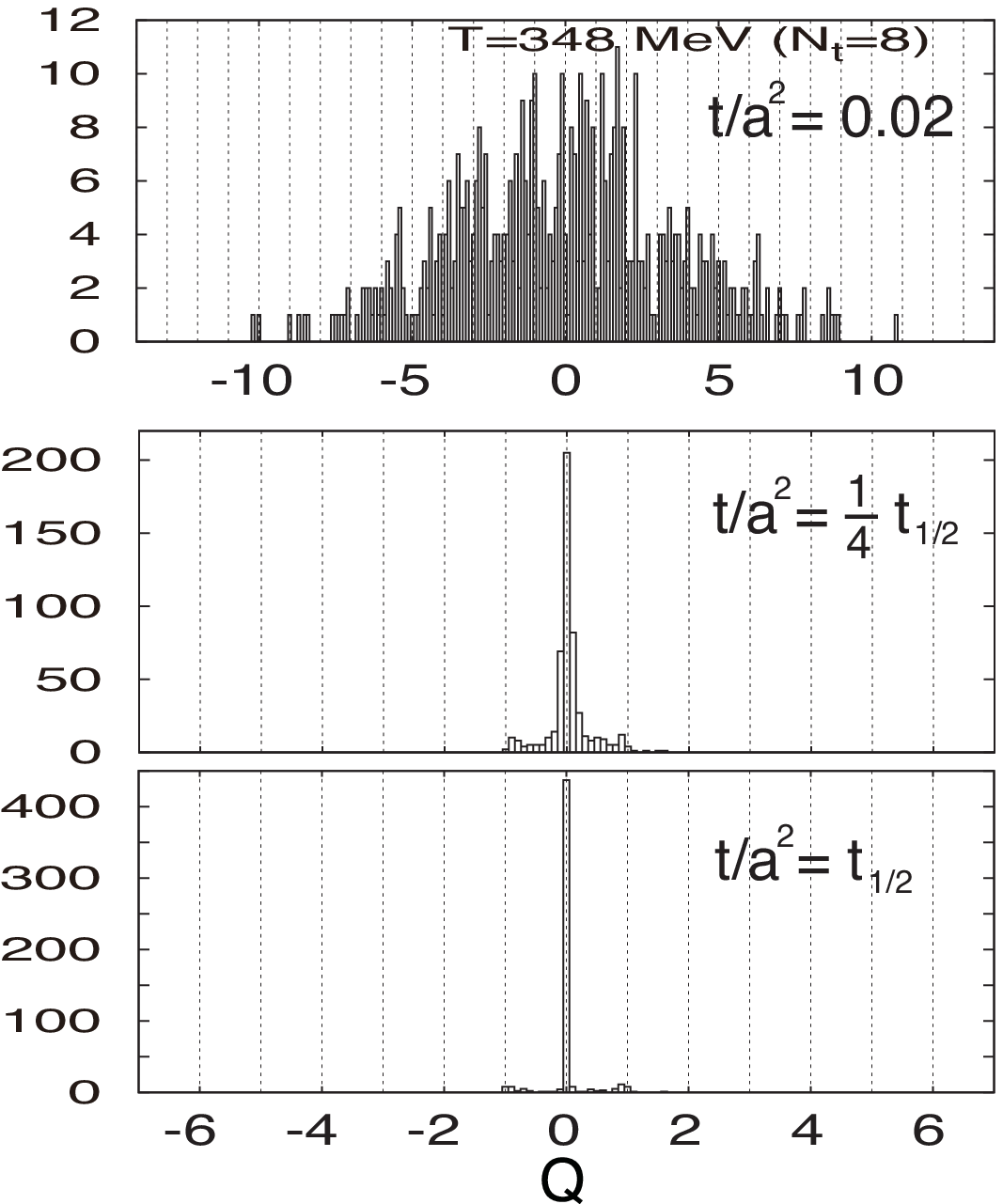}
 \end{center}
  \caption{Histogram of the topological charge with the gluonic definition at $T\simeq232$ MeV (left panel) and $T\simeq348$ MeV (right panel). The upper, middle, and lower plots in each panel are for flow times $t/a^2=0.02$, $\frac{1}{4} t_{1/2}$, and $t_{1/2}$, respectively.
}
\label{fig:histogram}
\end{figure}
In the left panel of Fig.~\ref{fig:histogram}, we plot the histogram of $Q$ with the gluonic definition at $T\simeq232$ MeV, obtained at various flow times.
We see that $Q$ accumulates to integer values as we flow the gauge configuration, i.e., the gradient flow works well as a renormalization with canonical normalization.
We find that $Q$ has well wide distribution on nonzero values at $T\simle279$ MeV ($N_t \simge 10$) 
but starts to freeze at $Q=0$ at $T\simge348$ MeV ($N_t \simle 8$), as shown in the right panel of 
Fig.~\ref{fig:histogram}. 

\begin{figure}[h]
 \begin{center}
  \includegraphics[width=7cm]{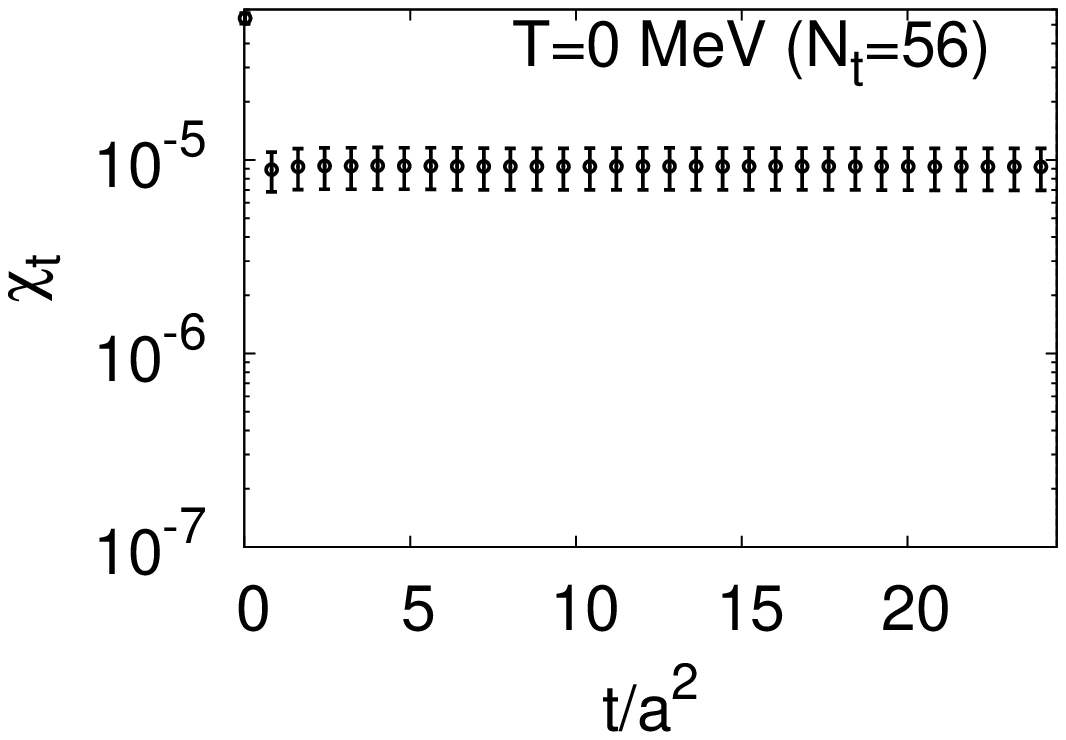}
  \includegraphics[width=7cm]{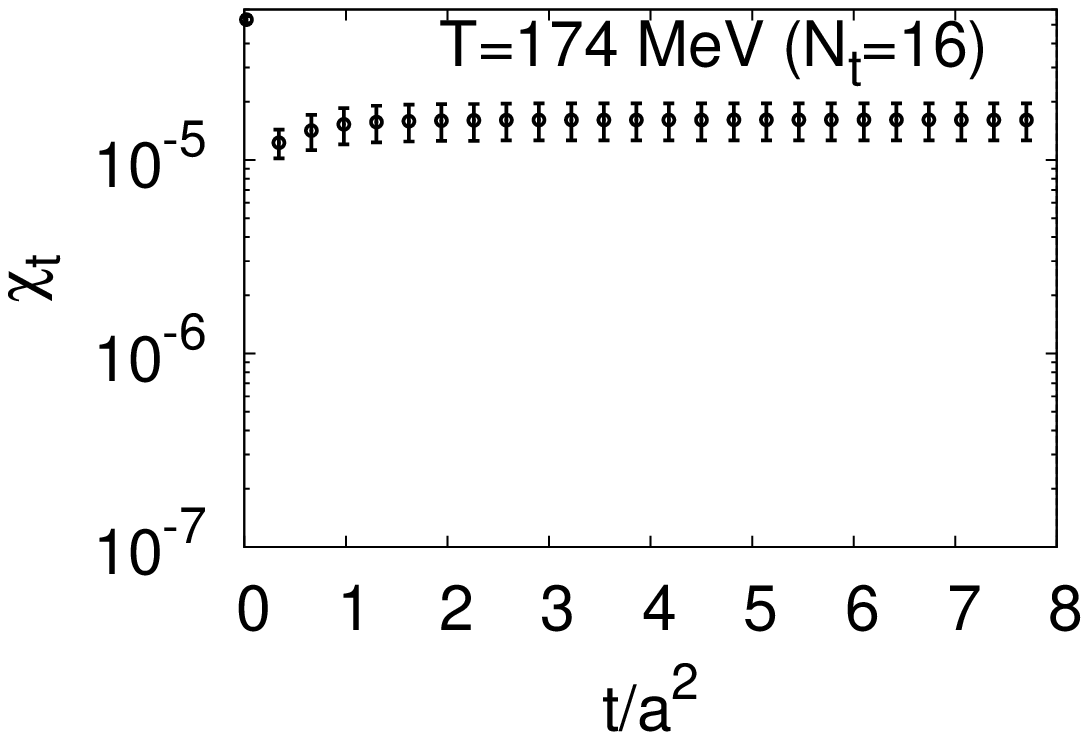}
  \includegraphics[width=7cm]{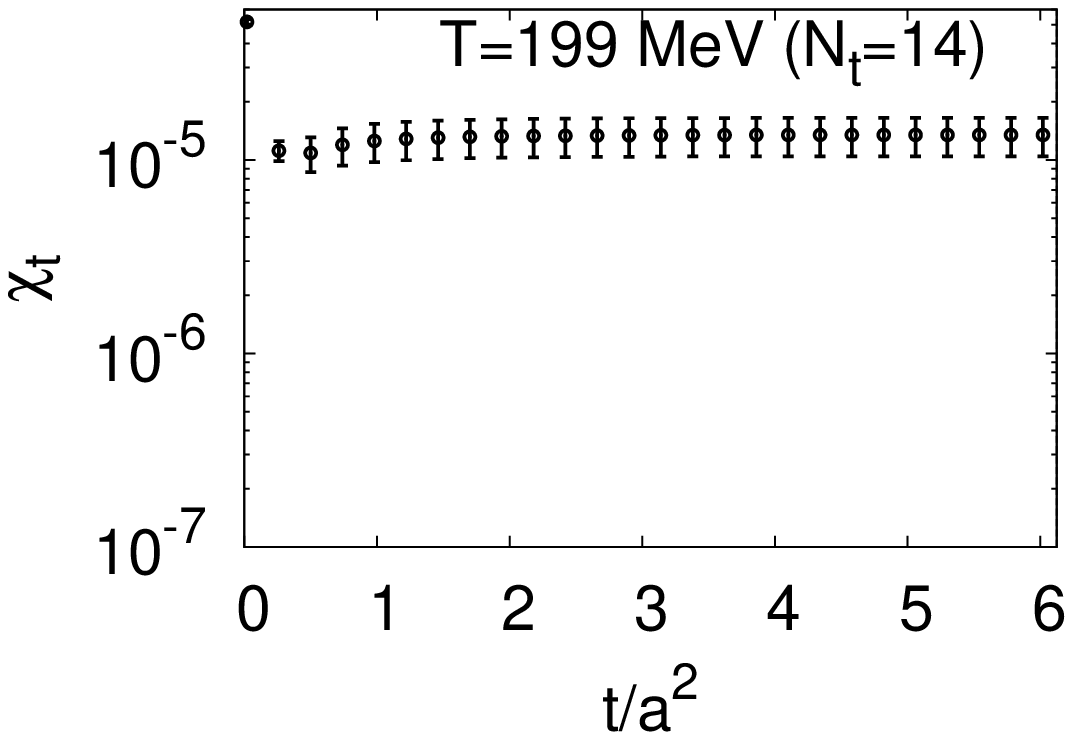}
  \includegraphics[width=7cm]{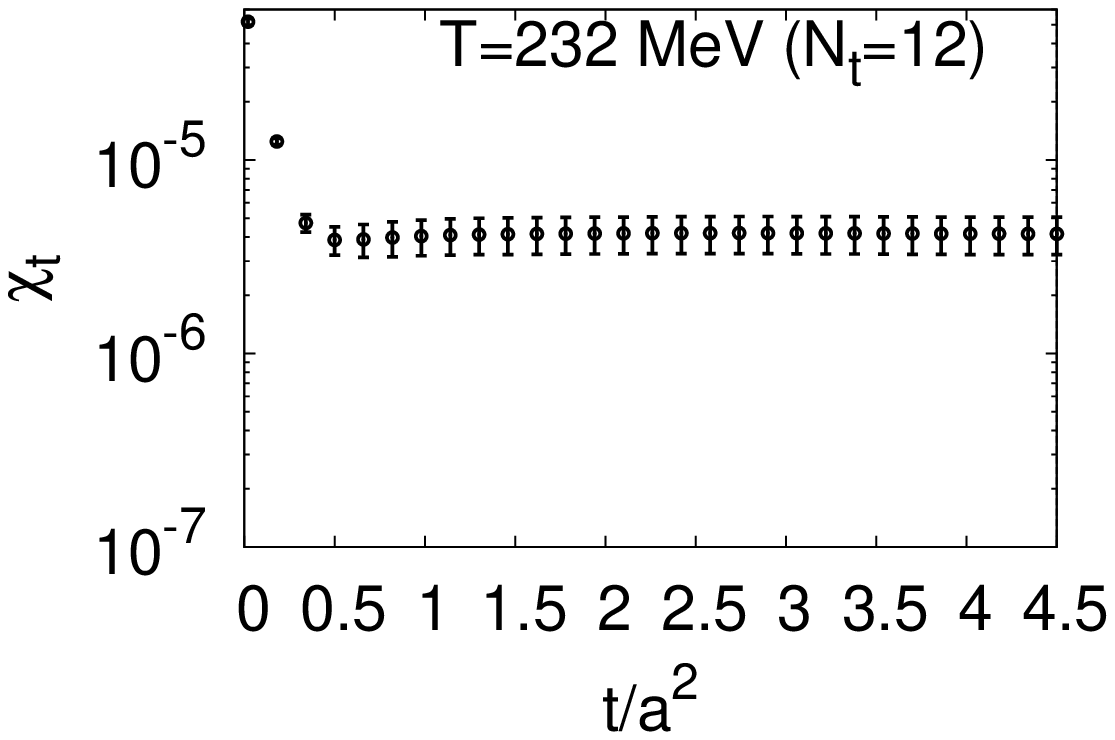}
  \includegraphics[width=7cm]{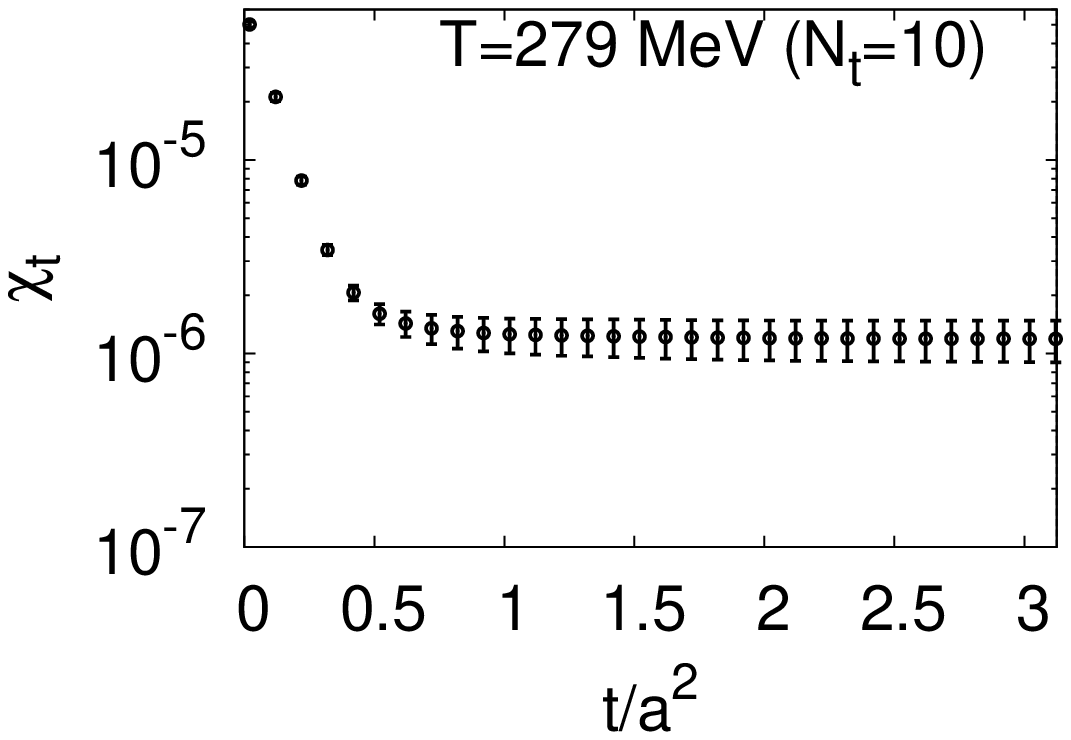}
  \includegraphics[width=7cm]{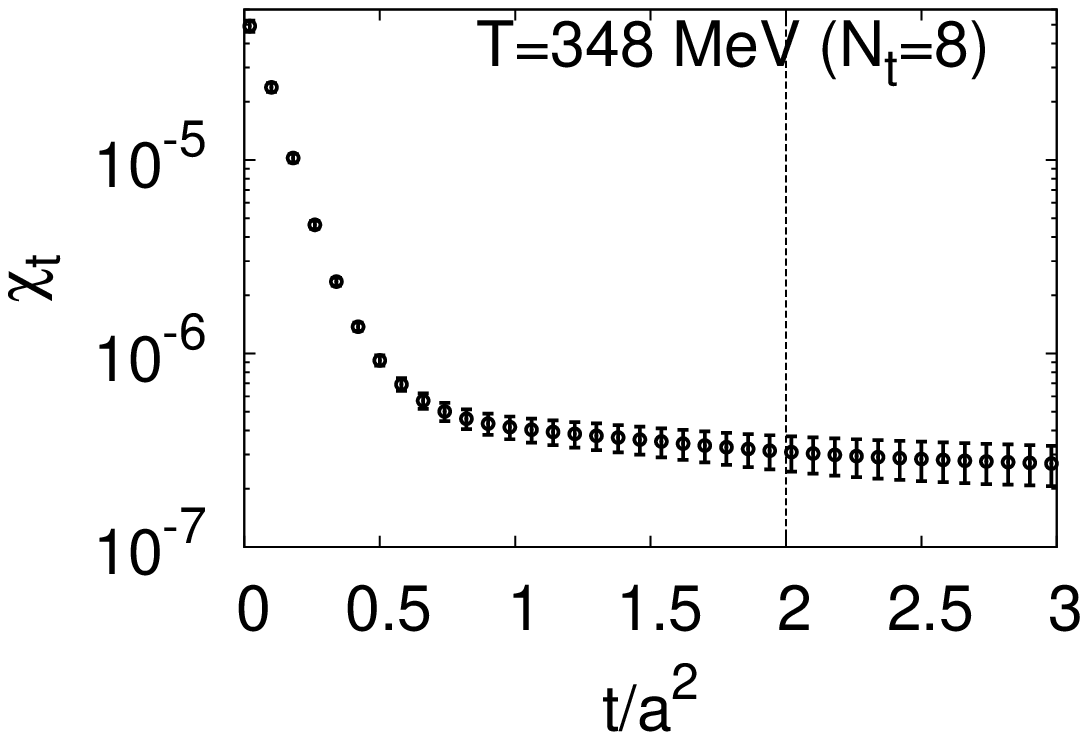}
  \includegraphics[width=7cm]{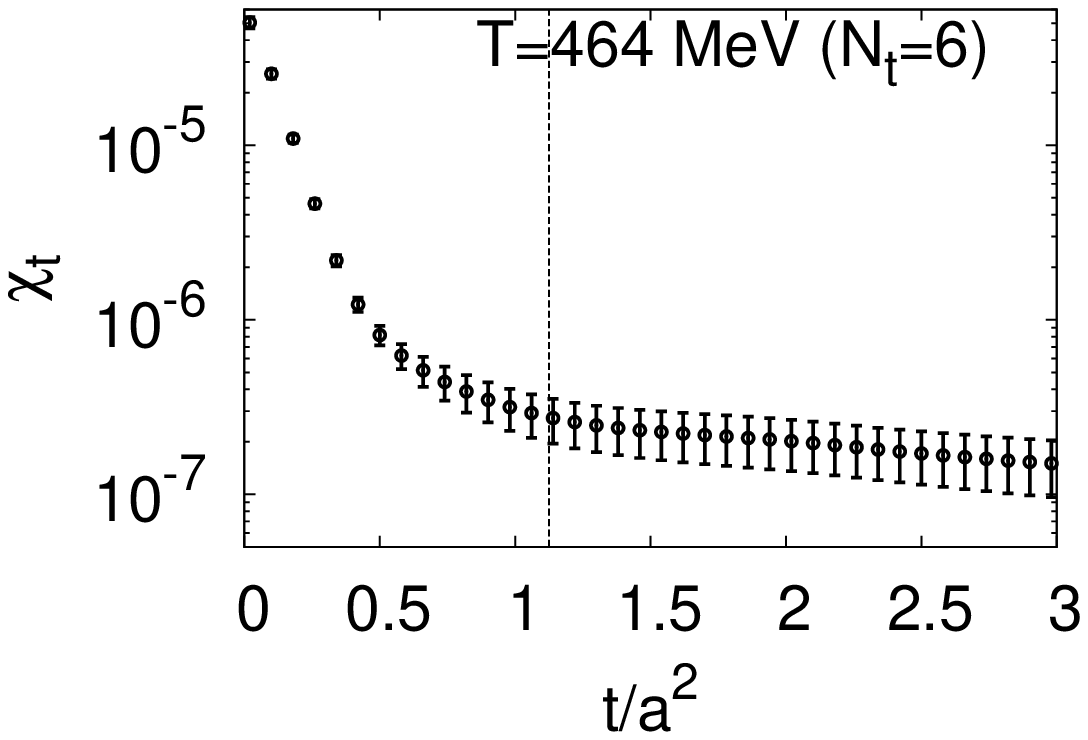}
  \includegraphics[width=7cm]{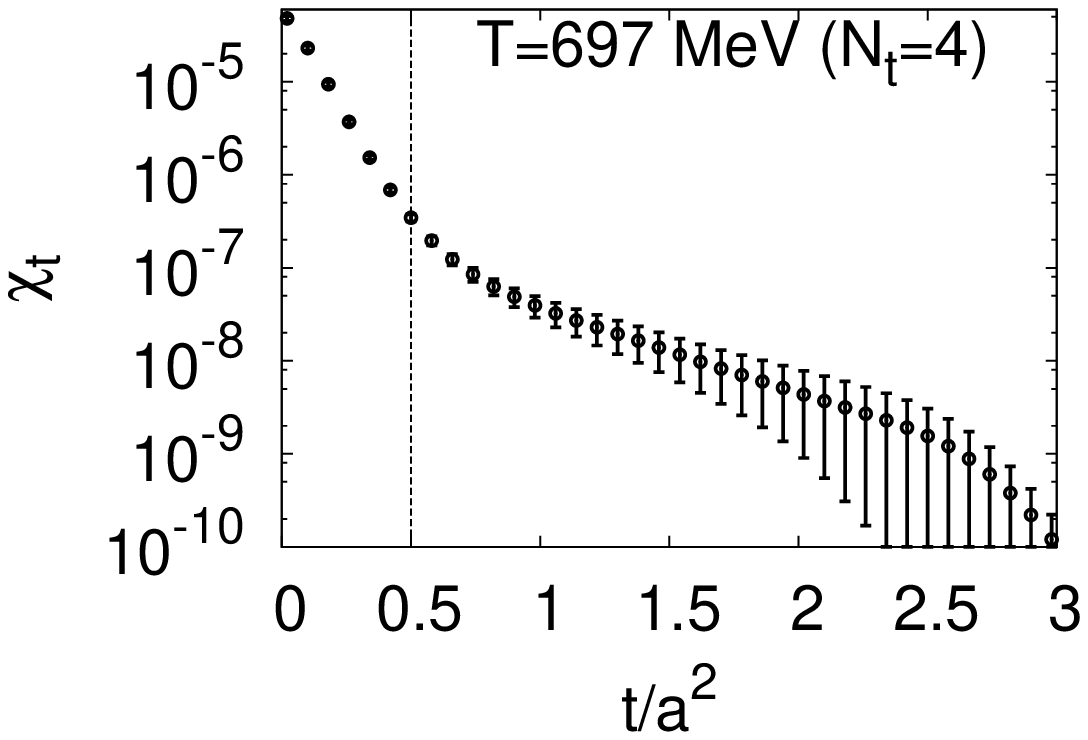}
 \end{center}
  \caption{The $\chi_\mathrm{t}(t,a)$ with the gluonic definition as a function of the flow time $t/a^2$.
From the top left: $T\simeq0$, $174$, $199$, $232$, $279$, $348$, $464$, and $697$ MeV, respectively.
Vertical axis is in lattice unit.
The vertical dotted lines for the higher three temperatures indicate $t_{1/2}$.
For the lower five temperatures, $t_{1/2}$ resides at the highest $t$ in the figure.
}
\label{eqn:Q_susceptibility_vs_t}
\end{figure}

The topological susceptibility with the gluonic definition is defined by
\begin{eqnarray}
 \chi_\mathrm{t}=\frac{1}{V_4}\left(\left\langle Q^2 \right\rangle
-\left\langle Q\right\rangle^2\right).
\label{eq:(1.5)}
\end{eqnarray}
In Fig.~\ref{eqn:Q_susceptibility_vs_t} we show the results of $\chi_\mathrm{t}(t,a)$ as a function of the flow time.
At $T\simle279$ MeV, we find wide plateaus below $t_{1/2}$ 
reflecting the flow time invariant property in the continuum~\cite{Luscher-talk,Ce:2015qha}.
On the other hand, at $T\simge348$ MeV, $\chi_\mathrm{t}(t,a)$ does not show a plateau up to~$t_{1/2}$.
On these lattices, we cannot extract a physical value due to lattice artifacts.
We thus concentrate on the range $T\simle279$ MeV. 
We also test other operators for the quadratic term of $G_{\mu\nu}$, including the clover operator only with plaquettes, that with only rectangle loops, and square of the imaginary part of the plaquette. We find that the results are consistent with each other within statistical errors. 
The results of the topological susceptibility with the gluonic definition are summarized later. 

\section{Fermionic definition}
\label{sec:fermionic}

In the continuum QCD, the topological susceptibility is related to the disconnected two point function of the flavor-singlet pseudoscalar
through chiral WT identities \cite{Bochicchio:1984hi,Giusti:2004qd},
\begin{eqnarray}
\left\langle \partial_\mu A_\mu^a(x) \, {\cal O}\right\rangle
-2m\left\langle \pi^a(x) \, {\cal O}\right\rangle
+2n_f\delta^{a0}\left\langle q(x) \, {\cal O}\right\rangle
=i\left\langle \delta^a{\cal O}\right\rangle,
\end{eqnarray}
where 
$n_{f}$ is the number of degenerate flavors with mass $m$ ($n_f=2$ and $m=m_{ud}$ in our case) and
$A_{\mu}^a(x)=\bar\psi(x)T^a\gamma_{\mu}\gamma_5\psi(x)$,
$\pi^a(x)=\bar\psi(x)T^a\gamma_5\psi(x)$
in which $T^a$ is the generator in the degenerate flavor space and $\psi$ is the multiplet of the degenerate flavors.
We set $T^0=1$ (i.e., $a=0$ stands for singlet) and
${\rm tr}\left(T^aT^b\right)=\delta^{ab}$ for $a,b\ge1$.

The desired relation is derived as follows:
From singlet WT identities for ${\cal O}=Q$ and ${\cal O}=P^{0}$,
\begin{eqnarray}
&&
-m\left\langle P^0 Q\right\rangle
+n_f\left\langle Q^2 \right\rangle
=0,
\\&&
-m\left\langle P^0 P^0\right\rangle
+n_f\left\langle Q \, P^0\right\rangle
=-\left\langle S^0\right\rangle,
\end{eqnarray}
where
$P^a\equiv\int d^4x \,\pi^a(x)$ and $S^a\equiv\int d^4x \,\bar\psi(x)T^a\psi(x)$, 
we obtain
\begin{eqnarray}
&&
n_f^2\left\langle Q^2 \right\rangle
=m^2\left\langle P^0 P^0\right\rangle-m\left\langle S^0\right\rangle .
\end{eqnarray}
On the other hand, 
for nonsinglet ${\cal O}=P^{b}$, 
\begin{eqnarray}
-2m\left\langle P^a P^b\right\rangle
=-\left(\delta^{ab}\frac{2}{n_f}\left\langle S^0\right\rangle
+d_{abc}\left\langle S^c\right\rangle
\right),
\end{eqnarray}
where $a,b,c\ge1$.
Since the nonsinglet flavor symmetry is not broken we get
\begin{eqnarray}
\chi_\mathrm{t}=\frac{1}{V_4}\left\langle Q^2 \right\rangle
=\frac{m^2}{V_4 n_f^2}\left(
\left\langle P^0 P^0\right\rangle-n_f\left\langle P^a P^a\right\rangle
\right),
\label{eqn:disconnected}
\end{eqnarray}
where the sum is not taken over $a$ in the right-hand side.
The right-hand side is nothing but the disconnected part of the
singlet pseudoscalar two point function.
The right-hand side of \eqn{eqn:disconnected} may have power divergences with Wilson or staggered fermions since the chiral symmetry is broken explicitly~\cite{Giusti:2004qd,Luscher:2004fu}.

To overcome the difficulties in the calculation of renormalized fermion bilinear operators due to violation of symmetries on the lattice, 
we adopt the method of Ref.~\cite{Hieda:2016lly} based on a small-$t$ expansion of gradient flow~\cite{Luscher:2011bx}.
The renormalized pseudoscalar density which satisfy the chiral WT identity is given by 
\begin{eqnarray}
m_{R}\left(\overline{\psi}_{f}\gamma_{5}\psi_{f}\right)_{R}
&=& \lim_{t\to0}
c_{S}(t) \, \Bar{m}_{\overline{\rm MS}}(1/\sqrt{8t})
\varphi_{f}(t)\,\overline{\chi}_{f}(t,x)\,\gamma_{5}\,\chi_{f}(t,x),
\label{eqn:cpp}
\end{eqnarray}
where
\begin{equation}
c_{S}(t)\equiv1+\frac{\Bar{g}_{\overline{\rm MS}}(1/\sqrt{8t})^2}{(4\pi)^2}
   \left[4\,(\gamma-2\ln2)
   +8
   +\frac{4}{3}\,\ln(432)\right]
\label{eqn:cs}
\end{equation}
is the matching factor between the gradient flow renormalization scheme and the ${\overline{\rm MS}}$ scheme~\cite{Hieda:2016lly},
and
\begin{equation}
   \varphi_f(t)\equiv
   \frac{-6}
   {(4\pi)^2\,t^2
   \left\langle\Bar{\chi}_f(t,x)\overleftrightarrow{\Slash{D}}\chi_f(t,x)
   \right\rangle_{\! 0}},
\quad
   \overleftrightarrow{D}_\mu\equiv D_\mu-\overleftarrow{D}_\mu ,
\label{eq:(1.18)}
\end{equation}
with $\langle \cdots \rangle_0$ the expectation value at $T=0$, is for the renormalization of fermion fields~\cite{Makino:2014taa}.
Here, $\Bar{g}_{\overline{\rm MS}}$ and $\Bar{m}_{\overline{\rm MS}}$ are the running coupling and mass
in ${\overline{\rm MS}}$ scheme.
Note that the combination in the left-hand side of \eqn{eqn:cpp} is independent of the renormalization scale.

With the fermionic definition, the extrapolation $t\to0$ is needed to remove contamination of unwanted dimension six operators after taking the continuum limit $a\to0$.
In numerical simulations, however, it is sometimes favorable to take the continuum extrapolation at a later stage.
At $a\ne0$, we have additional contaminations.
Since we adopt the nonperturbatively $O(a)$-improved Wilson fermion,
the lattice artifacts start with $O(a^2)$.
To the lowest orders of $a^2$, we expect 
\begin{eqnarray}
\chi_\mathrm{t}(t,a)&=&
\chi_\mathrm{t}+A\frac{a^2}{t}+tS
+\sum_{f}B_f(am_f)^2+C(aT)^2
+D\left(a\Lambda_{\rm QCD}\right)^2
+a^2S'+O(a^4,t^2),
\nn\\&&
\label{eqn:expansion}
\end{eqnarray}
where $\chi_\mathrm{t}$ in the right-hand side is the physical topological susceptibility,
$A$, $B$, $C$, $D$ are contributions from dimension four operators and
$S$, $S'$ are those from dimension six operators.
To exchange the limiting procedures $t\to0$ and $a\to0$, we need to remove the singular terms at $t=0$.
This may be possible if we can identify a ``window'' in $t$ where $\chi_\mathrm{t}(t,a)$ is dominated by the linear term of \eqn{eqn:expansion}.
In Ref.~\cite{ourEOSpaper}, we found that the energy-momentum tensor and the chiral condensate similarly computed on the same configurations do have clear linear windows when $t_{1/2}$ is not very small.

\begin{figure}[h]
 \begin{center}
  \includegraphics[width=7cm]{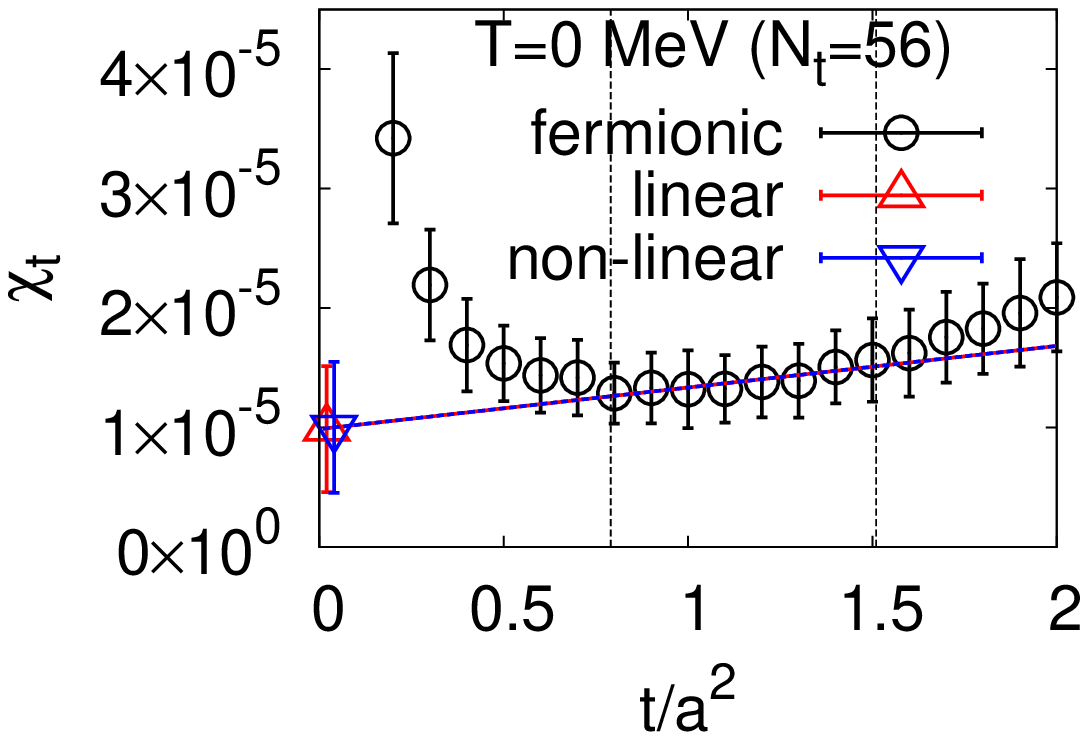}
  \includegraphics[width=7cm]{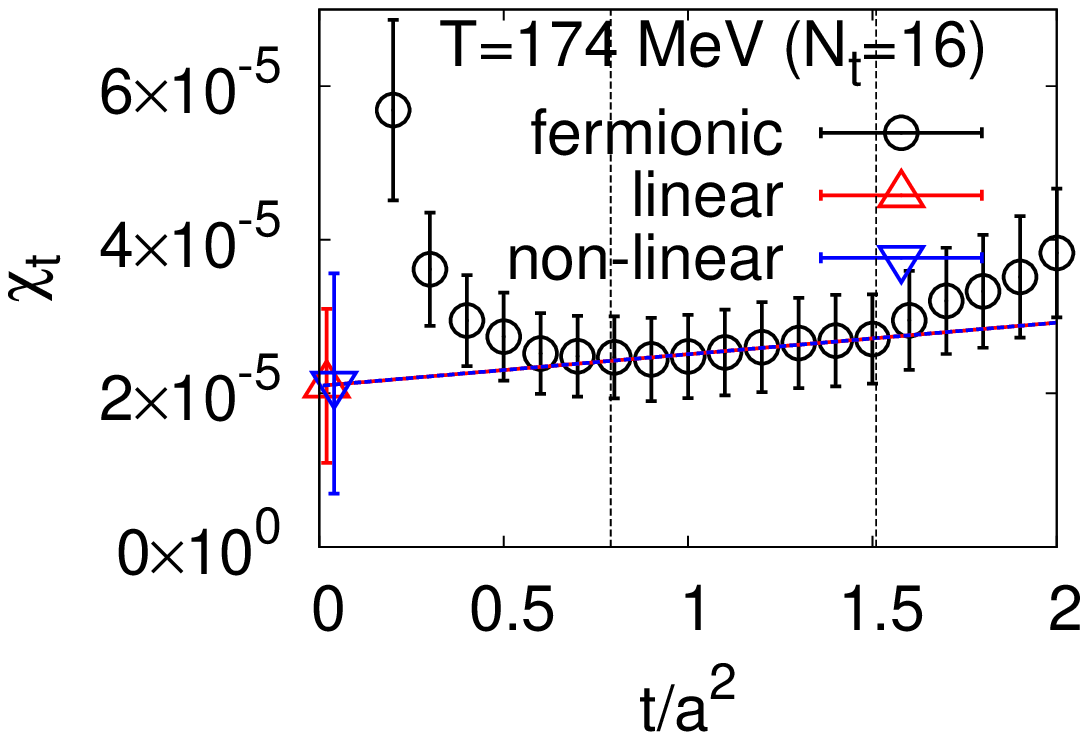}
  \includegraphics[width=7cm]{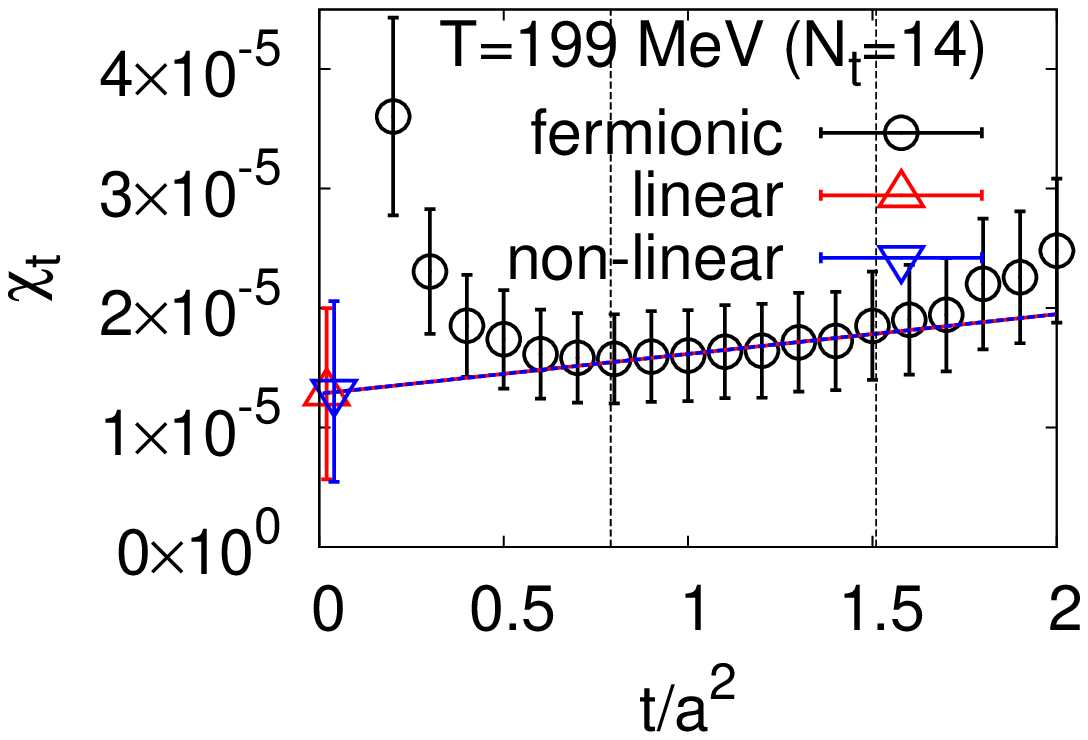}
  \includegraphics[width=7cm]{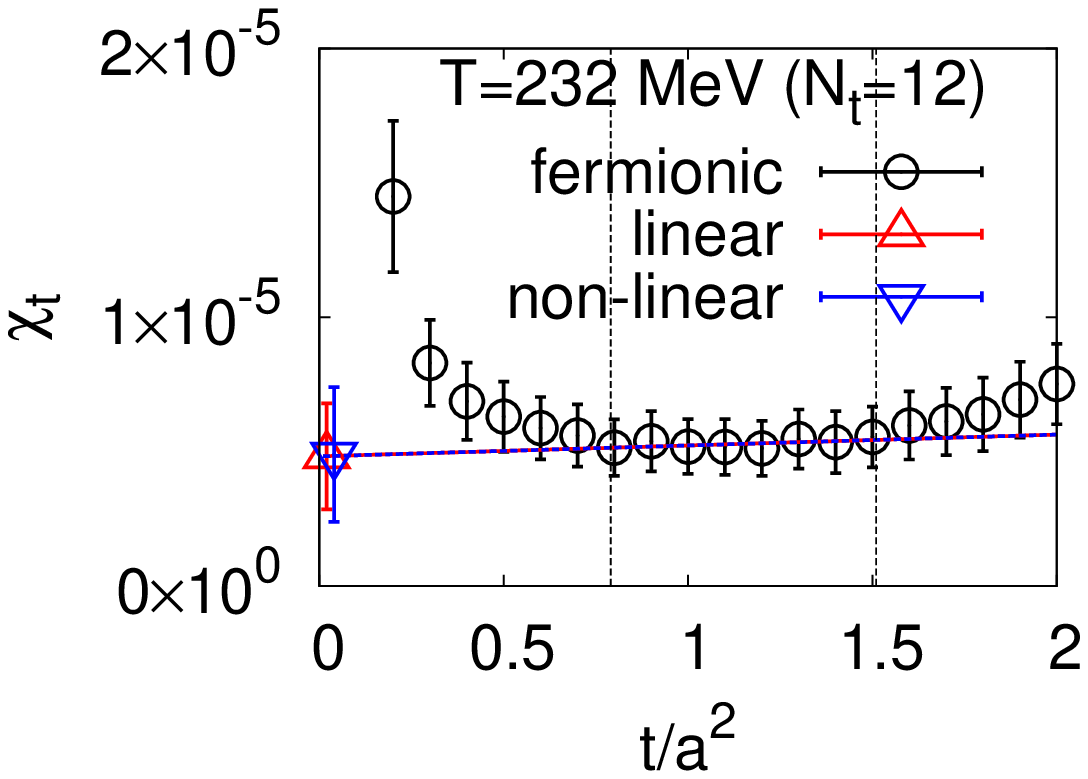}
  \includegraphics[width=7cm]{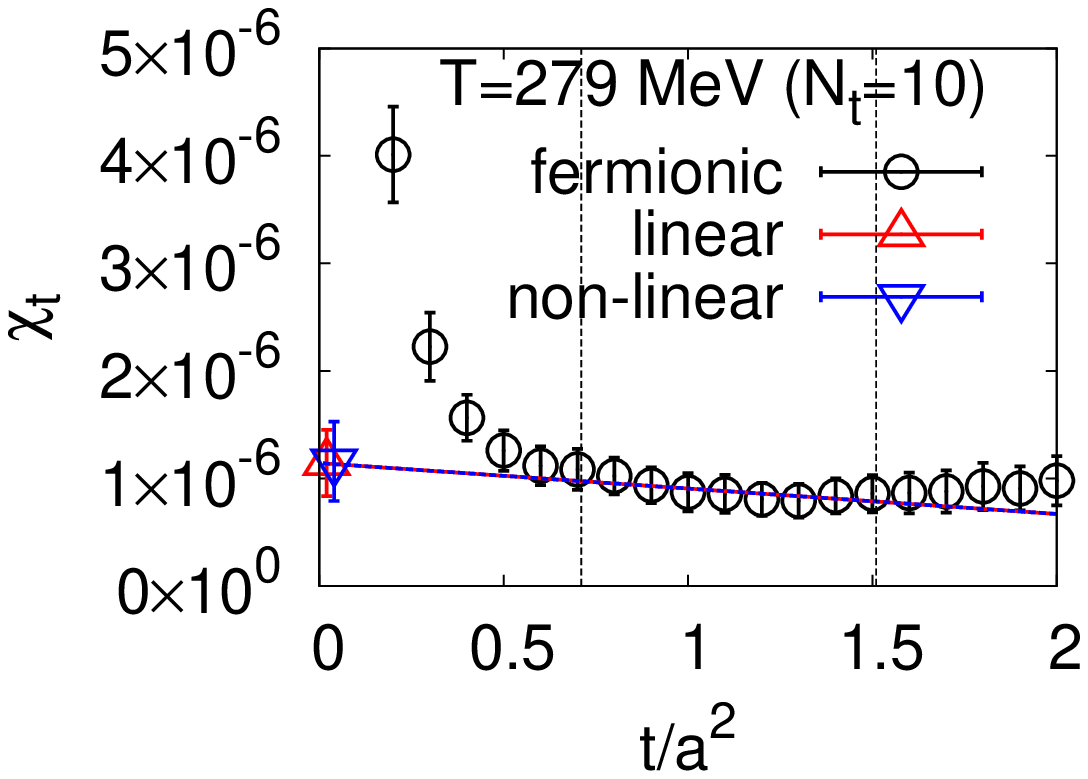}
  \includegraphics[width=7cm]{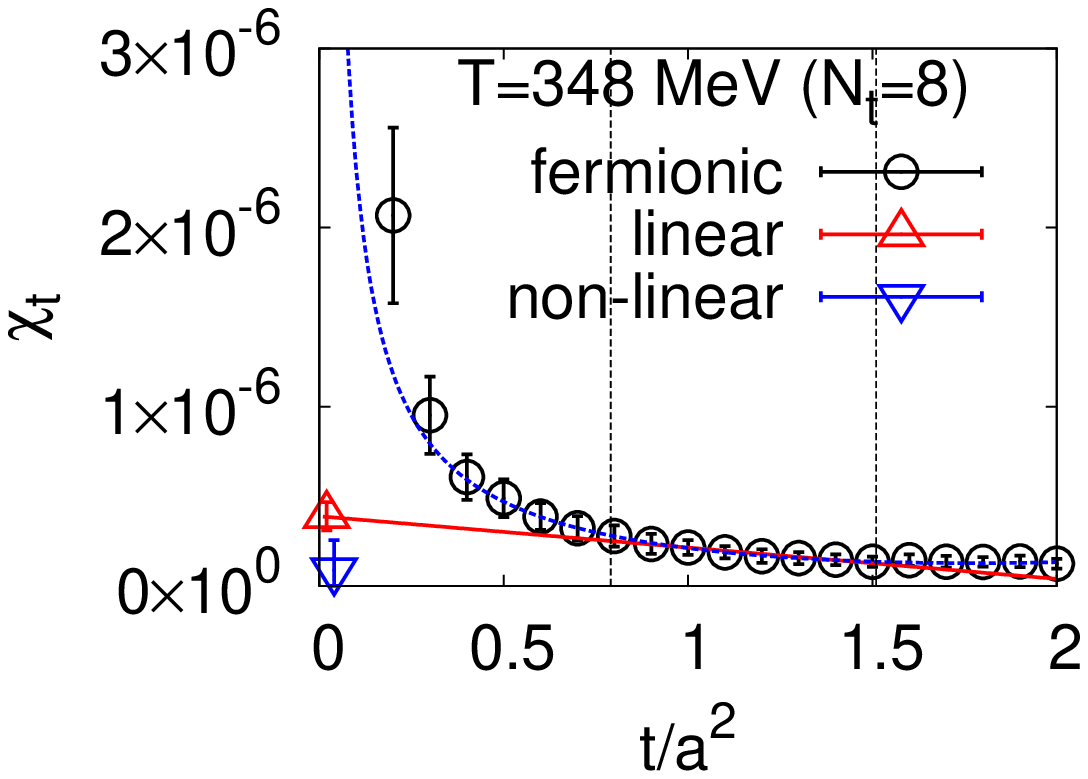}
  \includegraphics[width=7cm]{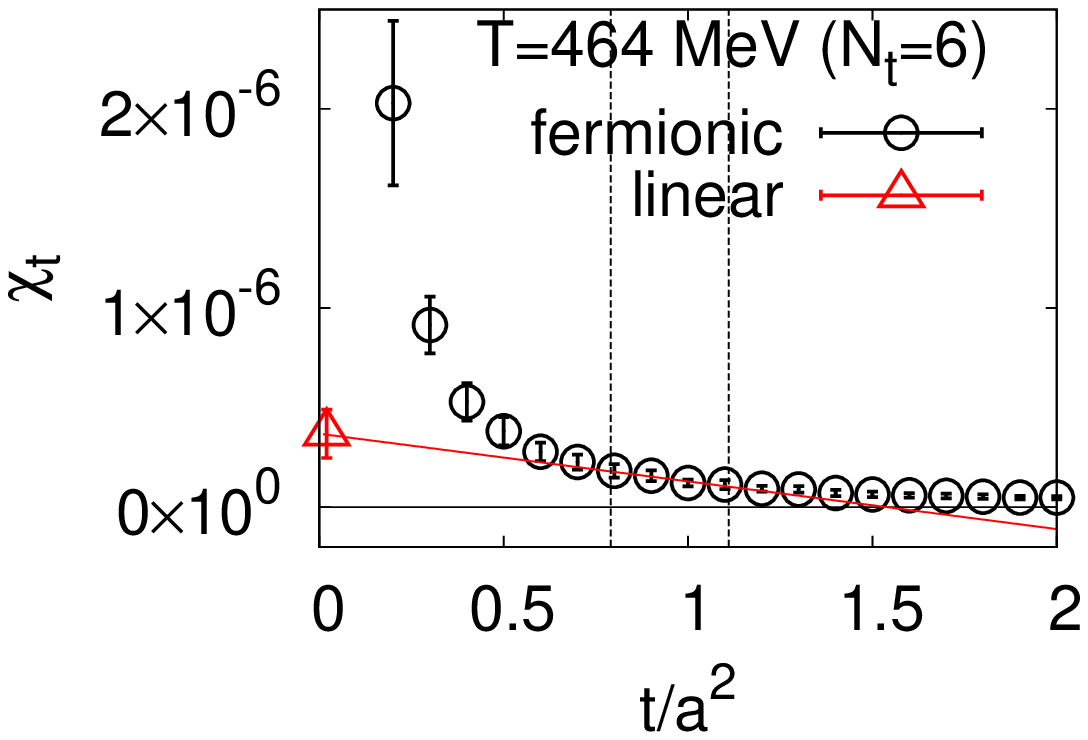}
  \includegraphics[width=7cm]{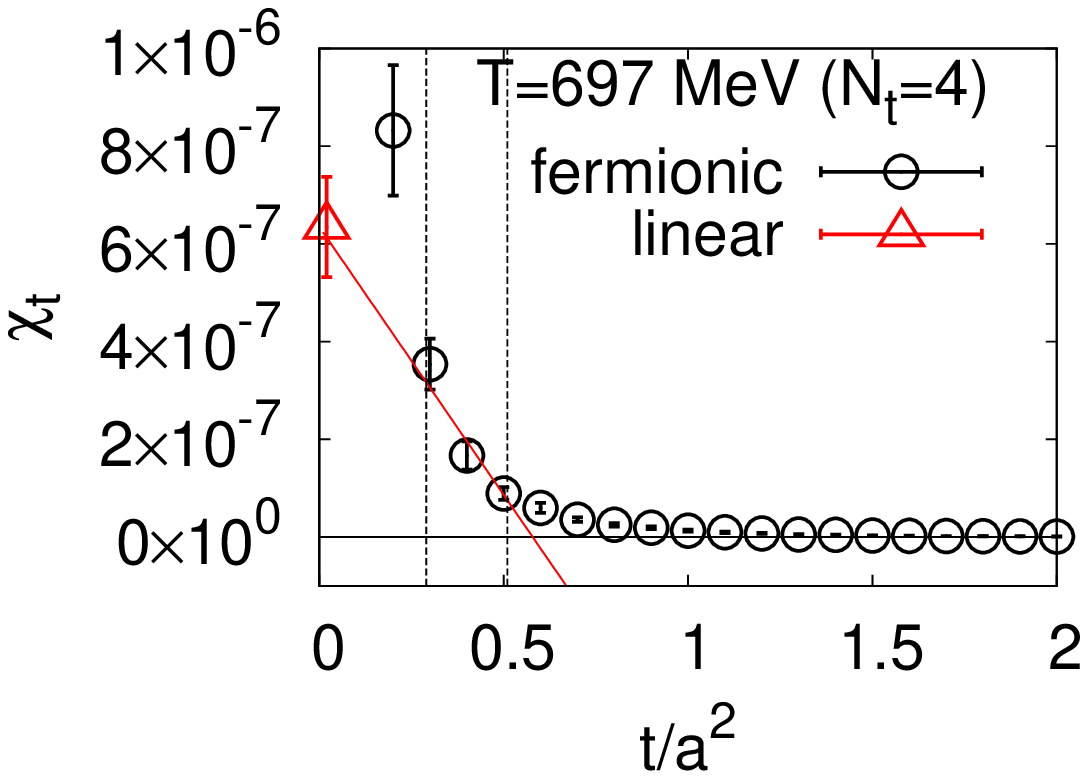}
 \end{center}
  \caption{$\chi_\mathrm{t}(t,a)$ with the fermionic definition as a function of the flow time $t/a^2$.
  From the top left: $T\simeq0$, $174$, $199$, $232$, $279$, $348$, $464$, and $697$ MeV, respectively.
  Vertical axis is in lattice unit.  Pairs of vertical dotted lines indicate
  the window for the fit.
  The red solid lines with upward triangles are the results of the linear fit.
  The blue dotted lines with downward triangles are the results of the nonlinear fit discussed in the text.
}
\label{fig:PP}
\end{figure}

In Fig.~\ref{fig:PP}, we plot $\chi_\mathrm{t}(t,a)$ for degenerate $u$ and $d$ quarks as a function of the flow time.
The nonlinear behavior near the origin may be due to the lattice artifact $a^2/t$ and that at large flow time due to the $O(t^2)$ contributions. 
At intermediate values of $t/a^2$, we find sufficiently wide liner windows well below $t_{1/2}$
for $T\simle348$ MeV.
On the other hand, for $T\simge 464$ MeV ($N_t\le6$) we could not identify a clear window below $t_{1/2}$ from our data. 
This will be in part due to the small $t_{1/2}$ on these lattices ($t_{1/2}=1.125$ and 0.5 for $N_t=6$ and 4, respectively).
Following the strategy of Ref.~\cite{ourEOSpaper}, we take the $t\to0$ limit by a linear fit using the data within the window for $T\simle348$ MeV. 
Results of the linear fits are shown by red solid lines with upward triangles in Fig.~\ref{fig:PP}.
For $T\simge 464$, we do make trial linear fits assuming linear windows with $t_{1/2}$ as the upper bounds, as shown in Fig.~\ref{fig:PP}, but the results should be treated with care because the windows are narrow.

In order to check the validity of the linear window and to estimate a systematic error due to the fit ansatz, 
we also try a nonlinear fit of the form
\begin{eqnarray}
\chi_\mathrm{t}(t,a)&=&
\chi_\mathrm{t}+A\frac{a^2}{t}+tB+Ct^{2}
\label{eqn:nonlinear-fit}
\end{eqnarray}
adopting the same windows for the fit range.
We restrict $A\ge0$ and $C\ge0$ to reproduce the increasing behavior of the data at small and large $t/a^2$ as seen in Fig.~\ref{fig:PP}.
The results of the nonlinear fit are shown by blue dotted lines with downward triangles in Fig.~\ref{fig:PP}.
We find that the coefficients $A$ and $C$ are very small and consistent with zero for
$T\simle279$ MeV, confirming the validity of the linear fit using the linear window.
At $T\simeq348$ MeV, on the other hand, we find a slight deviation from the linear fit, which we take as a systematic error in our final result.
For $T\simge464$, the nonlinear fit is not applicable because the number of data points within the window is not sufficient.
Because the linear fits are also not reliable for these temperatures, we just disregard the results at $T\simge464$.

\begin{figure}[h]
 \begin{center}
  \includegraphics[width=10cm]{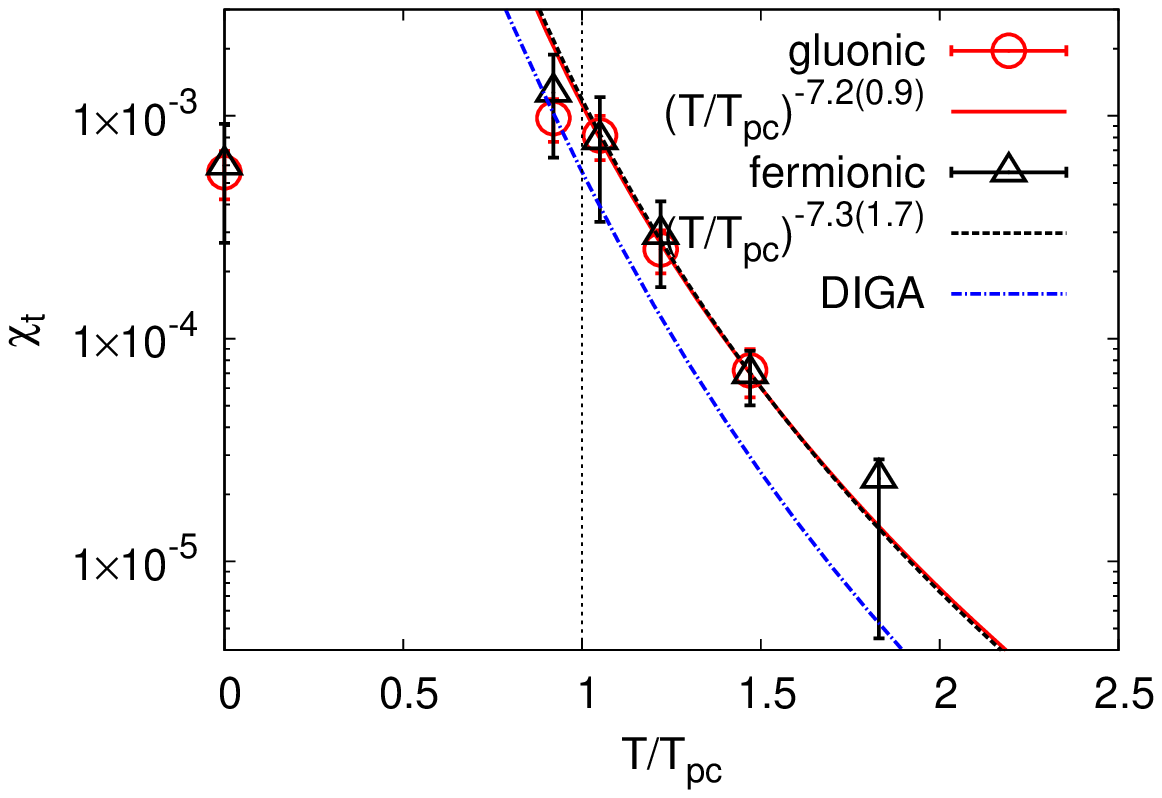}
 \end{center}
  \caption{Topological susceptibility as a function of temperature.
  Vertical axis is in unit of $({\rm GeV})^4$. 
  }
\label{fig:Q_susceptibility_gauge_vs_fermion}
\end{figure}

Our results of $\chi_\mathrm{t}$ with the gluonic and fermionic
definitions are summarized in
Fig.~\ref{fig:Q_susceptibility_gauge_vs_fermion} and Table~\ref{table:results}.
We find that the results from both definitions agree well with each other for~$T\simle279$ MeV ($T/T_\mathrm{pc}\simle1.47$). 

\begin{table}[htb]
\centering
\begin{tabular}{cccc}
 $T$ (MeV) & $T/T_{\mathrm{pc}}$ & $\chi_\mathrm{t}^{\rm gluonic}$ & $\chi_\mathrm{t}^{\rm fermionic}$  \\
\hline
 $0$   & $0$    & $0.00056(14)$  & $0.00060(32)(^{+3}_{-4})(0)$ \\
 $174$ & $0.92$ & $0.00098(21)$  & $0.00127(61)(^{+7}_{-14})(0)$ \\
 $199$ & $1.05$ & $0.00082(18)$  & $0.00078(43)(^{+4}_{-9})(0)$ \\
 $232$ & $1.22$ & $0.000251(55)$ & $0.00029(12)(^{+1}_{-2})(0)$ \\
 $279$ & $1.47$ & $0.000072(18)$ & $0.000069(19)(^{+3}_{-4})(0)$ \\
 $348$ & $1.83$ & NA & $0.0000235(47)(21)(^{+0}_{-183})$ \\
 $464$ & $2.44$ & NA & NA \\
 $697$ & $3.67$ & NA & NA \\
\hline
\end{tabular}
\caption{The topological susceptibility $\chi_\mathrm{t}$ with the
 gluonic and fermionic definitions.
 The unit is $({\rm GeV})^4$.
 The number in the first parenthesis is the statistical error estimated by a jackknife method.
 For the fermionic definition, the second parenthesis is the
 systematic error due to the perturbative coefficients \eqn{eqn:cs}
 and the running mass in \eqn{eqn:cpp}, 
 and the third parenthesis is the systematic error from the fit ansatz estimated by the difference between the linear and nonlinear fits. 
}
\label{table:results}
\end{table}%

Finally, we fit the data of $\chi_\textrm{t}$ at $T/T_\mathrm{pc}\simeq 1.05$--$1.47$ with a power low $(T/T_\mathrm{pc})^{-\gamma}$. 
The results are shown by red solid and black dashed curves for gluonic and fermionic definitions, respectively in Fig.~\ref{fig:Q_susceptibility_gauge_vs_fermion}.
For the exponent, we find $\gamma=7.2(0.9)$ and $7.3(1.7)$ with the gluonic and fermionic definitions, respectively.
These numbers are perfectly consistent with the one-loop DIGA prediction
$\gamma\simeq7.5$ at $T/T_\mathrm{pc}\simeq1.5$.
They agree even with the DIGA prediction $\gamma=8$ in the high temperature
limit within statistical errors. 
On the other hand, the absolute value of $\chi_\mathrm{t}$ is slightly larger than a prediction of DIGA 
(the blue dot-dashed curve in Fig.~\ref{fig:Q_susceptibility_gauge_vs_fermion}).
The DIGA prediction is given by an integration over the instanton size
$\rho$
\begin{eqnarray}
&&
\chi_\mathrm{t}(T)=2\int_0^\infty\frac{d\rho}{\rho^5} \,
n_G(\rho) \, n_F(m_{ud}\rho) \, n_T(\pi\rho T)
\end{eqnarray}
where $n_G$, $n_F$, and $n_T$ are the gauge, fermion, and finite temperature contributions, respectively, whose explicit forms are given by an instanton calculation \cite{tHooft:1976snw,Gross:1980br} and are summarized in Ref.~\cite{frison}.
Inputs for the calculation are the pseudocritical temperature
$T_\mathrm{pc}=190$ MeV, the QCD scale
$\Lambda_\textrm{QCD}^{\ovl{\rm MS}}=332(19)$ MeV \cite{Agashe:2014kda}
and the bare quark mass for which we adopted our PCAC mass of the
up and down quarks.
The $\overline{\rm MS}$ scheme is used with the renormalization scale which we 
set to $\mu=2\pi T$.

\section{Conclusion and discussion}
\label{sec:conclusion}

We study temperature dependence of the topological susceptibility with the gradient flow method in (2+1)-flavor QCD with heavy $u$ and $d$ quarks, $m_\pi/m_\rho\simeq0.63$, at a single but fine lattice spacing $a\simeq0.07$ fm.
We find that the results with the gluonic and fermionic definitions agree well with each other for $T\simle 279$ MeV
even with the Wilson-type quarks
whose numerical cost is much less than the Ginsparg-Wilson lattice quarks.
Although the continuum extrapolation is not taken yet, the good agreement of different methods suggests that our lattices are already close to the continuum limit and the results are quantitatively reliable, in accordance with the observation of~Ref.~\cite{ourEOSpaper} based on the results of the equation of state and the chiral condensate with gradient flow.
At higher temperatures, we encountered several difficulties due to small $N_t$ and due to topological freezing. 
For the former, we need to decrease $a$. 
For the latter, a new idea such as the proposal of Ref.~\cite{frison} is needed.

Our topological susceptibility at $T/T_\mathrm{pc}\simeq 1.05$--$1.47$ show a power low behavior 
$\chi_\mathrm{t} \propto (T/T_\mathrm{pc})^{-\gamma}$ 
with exponent consistent with the prediction of the DIGA within statistical errors.
Here, we note that there are discrepancies among our results and previous results, 
such as $\gamma=5.98(12)$ and $7.84(88)$ of Ref.~\cite{Petreczky:2016vrs} for the gluonic and fermionic definitions, respectively, in a similar temperature region, obtained with improved staggered quarks after taking a continuum extrapolation.
To investigate a source of the discrepancy, we need to repeat the study at lighter quark mass and different lattice spacings.
Studies are going on at a $m_{ud}$ close to the physical point and at different lattice spacings.

\acknowledgments
We thank other members of the WHOT-QCD Collaboration for valuable discussions.
This work is in part supported by JSPS KAKENHI Grants No.\ 26400251, No.\ 15K05041 and No.\ 16H03982,
by the Large Scale Simulation Program of High Energy Accelerator
Research Organization (KEK) No.\ 14/15-23, No.~15/16-T06, No.~15/16-T-07, No.~15/16-25, and No.~16/17-05, 
and by Interdisciplinary Computational Science Program in CCS, University of
Tsukuba.
This work is in part based on Lattice QCD common code Bridge++ \cite{bridge}.


\end{document}